\documentclass[conference]{IEEEtran}
\IEEEoverridecommandlockouts
\usepackage{amsmath,amssymb,amsfonts}
\usepackage[noend]{algcompatible}
\usepackage{graphicx}
\usepackage{textcomp}
\usepackage{xcolor}
\def\BibTeX{{\rm B\kern-.05em{\sc i\kern-.025em b}\kern-.08em
    T\kern-.1667em\lower.7ex\hbox{E}\kern-.125emX}}
\usepackage[flushleft]{threeparttable}

\usepackage{graphicx}
\usepackage{xcolor}
    
\usepackage{hyperref}
\hypersetup{
    colorlinks=true,
    linkcolor=black,
    filecolor=black,      
    urlcolor=black,
    citecolor=black,
}
\usepackage{multirow}
\usepackage{chngcntr}
\usepackage{algorithm}
\usepackage{algpseudocode}
\algdef{SE}{Begin}{End}{\textbf{begin}}{\textbf{end}}
\algnewcommand\algorithmicforeach{\textbf{for each}}
\algdef{S}[FOR]{ForEach}[1]{\algorithmicforeach\ #1\ \algorithmicdo}

\usepackage{lipsum}
\newcommand\blfootnote[1]{%
  \begingroup
  \renewcommand\thefootnote{}\footnote{#1}%
  \addtocounter{footnote}{-1}%
  \endgroup
}

\usepackage[natbib=true, style=numeric, maxnames=99, sorting=none]{biblatex}
\begin{document}

\title{Nonparametric Estimation and Comparison of Distance Distributions from Censored Data
}

\author{
    \IEEEauthorblockN{Lucas H. McCabe}
    \IEEEauthorblockA{
        \textit{Department of Computer Science, George Washington University}, Washington, DC, USA\\
    }
    \IEEEauthorblockA{
        \textit{Logistics Management Institute}, Tysons, VA, USA\\
        lucasmccabe@gwu.edu
    }
}

\maketitle

\begin{abstract}
Transportation distance information is a powerful resource, but location records are often censored due to privacy concerns or regulatory mandates. We outline methods to approximate, sample from, and compare distributions of distances between censored location pairs, a task with applications to public health informatics, logistics, and more. We validate empirically via simulation and demonstrate applicability to practical geospatial data analysis tasks.
\end{abstract}

\begin{IEEEkeywords}
censored data analysis, transportation data, survival analysis, statistical testing, density estimation
\end{IEEEkeywords}

\section{Introduction}

\blfootnote{© 2024 IEEE.  Personal use of this material is permitted.  Permission from IEEE must be obtained for all other uses, in any current or future media, including reprinting/republishing this material for advertising or promotional purposes, creating new collective works, for resale or redistribution to servers or lists, or reuse of any copyrighted component of this work in other works.}

Analysis of transportation data is often constrained by the censoring of location records. This can occur in the public health domain, where precise location data may be risky to share. It has been reported that rural Americans live, on average, over twice as far from their nearest hospital as their counterparts in urban areas \cite{lam2018far}. What impact does this have on the practical experiences of patients in emergency situations? For instance, evidence suggests that incident location-to-destination ambulance journey distance is positively correlated with patient mortality \cite{nicholl2007relationship}. Such questions are important to the study of equity in medicine, but they can be more difficult to answer when location information is censored.

The study of migration flows can also be hindered by location-censoring. As part of the American Community Survey (ACS), the United States Census collects detailed migration information, but the data is censored (e.g., to county-level) for public release \cite{acs_web}. Franklin and Plane underscore this geographic censoring (e.g., varying spatial scales over time) as a limitation for migration research and express interest in more informative records while ensuring confidentiality \cite{franklin2006pandora}. 

In such situations, some researchers assign event origin and destination points to the centroid (or other representative point) of their corresponding censored regions \cite{rietveld1999relationship, boscoe2012nationwide}. The National Bureau of Economic Research hosts a County Distance Database, containing the Haversine distances between internal points of U.S. counties, which can support such analyses \cite{nber_county_distance}. When the level of spatial aggregation is relatively small for the problem, the resulting approximation can be satisfactory \cite{bliss2012estimating}, but this is not always the case and can lead to errors and contradictory results \cite{mizen2015quantifying}.

It is important to consider the informativeness of the censoring mechanism. For instance, in surveys where respondents offer intervals covering a quantity of interest, the intervals provide information about the true values \cite{angelov2017nonparametric}. In an estimation of length distributions from line segment processes, censoring arises from the sampling scheme, allowing authors Pawlas and Zikmundov\'{a} to treat the mechanism as uninformative \cite{pawlas2019comparison}. In our case, the censoring of transportation events to administrative boundaries is done to meet privacy standards, which we consider uninformative, as well.

In this work, our contributions are as follows:
\begin{itemize}
    \item We treat the estimation of transportation event distance distributions from location-censored records as a survival analysis problem (Section \ref{sec:method}).
    \item We present a stochastic dominance test for collections of location-censored transportation events (Section \ref{sec:distance}).
    \item We validate our approach empirically via simulation (Section \ref{sec:simdesign}, \ref{sec:simresults}) and demonstrate practical applicability by reanalyzing data from a study involving breast cancer screening (Section \ref{sec:maheswaran}).
\end{itemize}

Although we are, of course, not the first to employ survival analysis for transportation geography (e.g., \cite{kadokawa2019highway, massa2020distance}) nor the first to compare survival curves (e.g., \cite{mantel1966evaluation, dormuth2022test}), we present a framework for comparing collections of censored transportation events and provide a sampling-based U-test for the task.

\section{Transportation Event Distance Distribution Reconstruction}\label{sec:length_meth}

\subsection{Overview}

Briefly, we define:
\begin{itemize}
    \item a \textit{location} as a geographical point characterized by coordinates,
    \item a \textit{locale} as a censored representation of a location, and 
    \item a \textit{transportation event} as an origin-destination pair. When the origin and destination are locations, it is an \textit{uncensored} transportation event. When the origin and destination are locales, it is a \textit{censored} transportation event.
\end{itemize}

The distance associated with a transportation event is the distance between start and end locations, a lower-bound on the actual distance that may have been traveled. Provided a sample of \emph{censored} transportation events, we aim to estimate the distribution of the \emph{uncensored} transportation event distances.

\subsection{Drawing Uncensored Samples}\label{sec:method}

We are provided a list of censored transportation events, where each event is a tuple of origin and destination locales.

\begin{enumerate}
\item \textbf{Obtain Distance Intervals: } For each start and end locale, we extract coordinates along the boundary ($\{b_{\text{start}_1}, b_{\text{start}_2}, \ldots \}$ and $\{b_{\text{end}_1}, b_{\text{end}_2}, \ldots \}$, respectively), which may be obtained from administrative shape files.

For each event, we construct a distance matrix \(D\) with elements $D_{ij} = f(b_{\text{start}_i}, b_{\text{end}_j})$ and distance function $f$. The event's observation interval $[\min(D), \max(D)]$ includes the true distance of the transportation event.

\item \textbf{CDF Estimation: } Our distance intervals are interval-censored event records. We estimate the complementary cumulative distribution (survival function) using Turnbull's nonparametric estimator, provided by the iterative algorithm detailed in \cite{turnbull1976empirical}. For computational efficiency, the survival function may be obtained by an expectation-maximization algorithm, as provided in \cite{anderson2017efficient}.

For any $d$, we may construct an $100 (1 - \alpha)\%$ confidence interval $\Big(\hat{S}_{\text{lower}, \alpha}(d), \hat{S}_{\text{upper}, \alpha}(d) \Big)$ from our estimated survival function $\hat{S}$ via the exponential Greenwood formula, as provided by \cite{sawyer2003greenwood, kalbfleisch2011statistical}:
\begin{equation}
    \widehat{Var}[\hat{S}(d)] = \frac{1}{{\left(\log(\hat{S}(d))\right)}^2} \sum_{d_i \leq d} \frac{y_i}{n_i(n_i - y_i)}
\end{equation}
\begin{equation}
    g(d) = \log(-\log(\hat{S}(d)))
\end{equation}
\begin{equation}
    h(d) = Q\Big(\frac{\alpha}{2}\Big) \sqrt{\widehat{Var}[\hat{S}(d)]}
\end{equation}
\begin{equation}
\hat{S}_{\text{lower}, \alpha}(d), \hat{S}_{\text{upper}, \alpha}(d) = \exp(-\exp(g(d) \pm h(d)))
\end{equation}

where $Q$ is the quantile function of the standard normal distribution, $n_i$ is the number of events with distance at least $d_i$, and $y_i$ is the number of events with distance between $d_i$ and $d_{i+1}$. Obtaining the estimated cumulative distribution function (CDF) is straightforward via $\hat{F}(d) = 1 - \hat{S}(d)$, and the confidence intervals are:
\begin{equation}
    \hat{F}_{\text{lower}, \alpha}(d) = 1 - \hat{S}_{\text{upper}, \alpha}(d)
\end{equation}
\begin{equation}
    \hat{F}_{\text{upper}, \alpha}(d) = 1 - \hat{S}_{\text{lower}, \alpha}(d)
\end{equation}

\item \textbf{Survival Function Sampling: } We sample in two stages to account for the uncertainty in estimating $F$, beginning with inverse transform sampling:
\begin{equation}
d_* = \hat{F}^{-1}(u), \quad u \sim \text{Uniform}(0,1)\nonumber
\end{equation}

Assuming the uncertainty to be Gaussian, we draw:
\begin{equation}
    \epsilon \sim \mathcal{N}(0, \sigma^2), \quad \sigma = \frac{\hat{F}_{\text{upper}, \alpha}(d_*) - \hat{F}(d_*)}{Q(1-\frac{\alpha}{2})} \nonumber
\end{equation}
\begin{equation}
    d_{\text{sample}} = d_* + \epsilon\nonumber
\end{equation}
Where $d_{\text{sample}}$ is our sampled distance.
\end{enumerate}

\subsection{Stochastic Dominance}\label{sec:dominance}

To illustrate the application of this approach to practical data analysis workflows, we consider the following situation: we are provided collections $E_A =\{e_{A, 1}, e_{A, 2}, \ldots, e_{A, i}\}$ and $E_B =\{e_{B, 1}, e_{B, 2}, \ldots, e_{B, j}\}$, representing censored transportation events of types $A$ and $B$, respectively. We wish to test if transportation events of type $A$ are consistently longer than those of type $B$.

Because the sampling procedure is non-deterministic, instability of downstream test statistics and p-values may arise. This can be addressed by repeating a hypothesis test, collecting the p-values, and applying Fisher's combined probability test. Given p-values $p_1, \dots, p_k$, the combined test statistic is:
\begin{equation}
T = -2 \sum_{i=1}^k \log(p_i),
\end{equation}
where $T \sim \chi^2_{2k}$ under the null hypothesis of uniformly distributed and independent p-values \cite{fisherstatistical}.

Taken together, these steps may be organized into a Monte Carlo U-test (Algorithm \ref{alg:compare_events_repeated}): for a predetermined number of trials, we repeatedly draw uncensored distance samples by applying Steps 1-3 of Section \ref{sec:method} before performing a one-sided U-test \cite{mann1947test}. We aggregate the resultant p-values using Fisher's method.

\begin{algorithm}
\caption{Monte Carlo U-Test for Censored Events}
\begin{algorithmic}[1]
\Procedure{MCUTest}{$E_A, E_B, m, n, n_{\text{trials}}$}
    \State \textbf{Input:} $E_A, E_B$: samples of censored transportation events of types $A$ and $B$, respectively. $m, n$: numbers of samples to draw from $F_A$ and $F_B$, respectively. $n_{\text{trials}}$: number of trials to run the underlying U-test. 
    \State \textbf{Output:} Test statistic $t$. Aggregated p-value $p$.\\

    \State Estimate $\hat{F}_A(d)$ using $E_A$
    \State Estimate $\hat{F}_B(d)$ using $E_B$

    \State $t \gets 0$

    \For{$i = 1$ to $n_{\text{trials}}$}
        \State $s_A \gets$ $m$ samples drawn from $\hat{F}_A(d)$
        \State $s_B \gets$ $n$ samples drawn from $\hat{F}_B(d)$

        \State $S \gets$ sorted list of items from $s_A$ and $s_B$

        \State $R_1 \gets \sum_{x \in S_A} \Big(\text{rank of $x$ in $S$}\Big)$
        \State $u = R_1 - \frac{m(m+1)}{2}$
        \State $p_i \gets P(U \geq u | F_A = F_B)$

        \State $t \gets t + \log(p_i)$
    \EndFor

    \State $t \gets -2 t$
    
    \State $p = P(T \geq t | T \sim \chi^2_{2n_{\text{trials}}})$

    \Return $t, p$
\EndProcedure
\end{algorithmic}\label{alg:compare_events_repeated}
\end{algorithm}

\subsection{Statistical Distance}\label{sec:distance}

Given $E_A$ and $E_B$, we may wish to estimate some notion of distance between their uncensored generating distributions. One such notion is the Kolmogorov-Smirnov (KS) statistic. We may obtain $\hat{F}_A$ and $\hat{F}_B$ for collections $E_A$ and $E_B$, respectively, via Steps 1-2 in Section \ref{sec:method} before calculating \cite{an1933sulla}:
\begin{equation}
    K = \max_{d} | \hat{F}_{A}(d) - \hat{F}_{B}(d) |.
\end{equation}
This is equivalent to the approach described in \cite{white2021analysis}, which calculates the maximum vertical distance between patient survival curves.

Alternatively, we may consider Maximum Mean Discrepancy (MMD), a statistical distance between distributions expressed as the largest difference in expected values across functions of a class $\mathcal{G}$, described in detail in \cite{gretton2006kernel, JMLR:v13:gretton12a}. Absent direct access to their uncensored generating distributions, we can draw distance samples $s_A$ (of size $m$) and $s_B$ (of size $n$) from $\hat{F}_A$ and $\hat{F}_B$, respectively, and apply the empirical MMD estimate provided in \cite{gretton2006kernel}:
\begin{align}
    \text{MMD}_\mathcal{G}(s_A, s_B) = \bigg( &\frac{1}{m^2} \sum_{i,j=1}^{m} k(s_{A,i}, s_{A,j}) \\ \nonumber
    &+ \frac{1}{n^2} \sum_{i,j=1}^{n} k(s_{B,i}, s_{B,j}) \\ \nonumber
    &- \frac{2}{mn} \sum_{i=1}^{m} \sum_{j=1}^{n} k(s_{A,i}, s_{B,j}) \bigg)^{1/2}
\end{align}
with kernel $k$, which we take to be the radial basis function kernel with bandwidth parameter $\sigma$:
\begin{equation}
k(x, y) = \exp\left(-\frac{\|x - y\|^2}{2\sigma^2}\right)
\end{equation}

\section{Empirical Validation}\label{sec:val}

\subsection{Simulation Design}\label{sec:simdesign}

\begin{figure}[t]
    \centering
    \includegraphics[width=0.325\textwidth]{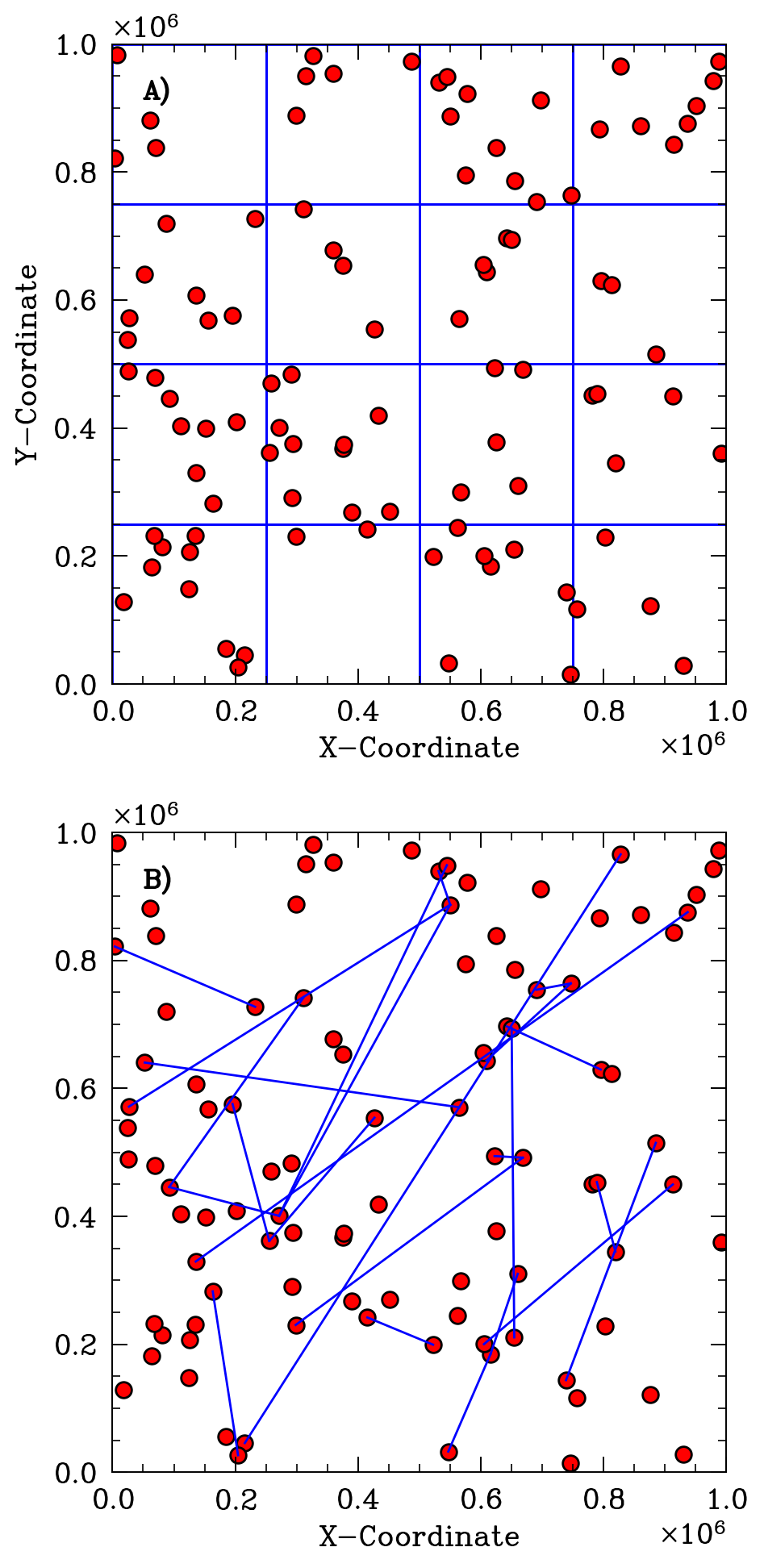}
    \caption{\textbf{A)} Visual example of the simulation environment, featuring a $1000000 \times 1000000$ grid with $100$ locations and $16$ locales. \textbf{B)} Visualization of resultant spatial network after generating $25$ transportation events.}
    \label{fig:gridworld_example}
\end{figure}

We conduct a brief simulation study to evaluate the performance of the proposed approach. We design a simple two-dimensional environment divided into equally-sized rectangular locales, with locations determined uniformly at random. Transportation event generation simulates trips between locations on the grid. Source locations are selected uniformly at random, and the destination of each trip is chosen based on an inverse distance policy, where the probability of choosing a location as the destination is inversely proportional to its distance $f$ from the source:
\begin{align}
P(\text{dest.} &= y_i | \text{source} = x) =
\frac{1}{(1+ f(x, y_i)) \sum_{j \neq i} \frac{1}{1 + f(x, y_j)}}
\end{align}
for $y_i \neq x$, otherwise the probability is assigned zero.

We map locations back to their corresponding locales to obtain censored and uncensored realizations of each event generated. An example environment is illustrated in Figure \ref{fig:gridworld_example}A. Transportation events can be interpreted as a spatial network, where nodes represent locations, edges represent transportation events, and edge weights convey the frequency of their corresponding transportation events (Figure \ref{fig:gridworld_example}B.)

\subsection{Simulation Results}\label{sec:simresults}

\begin{figure}[t]
    \centering
    \includegraphics[width=0.325\textwidth]{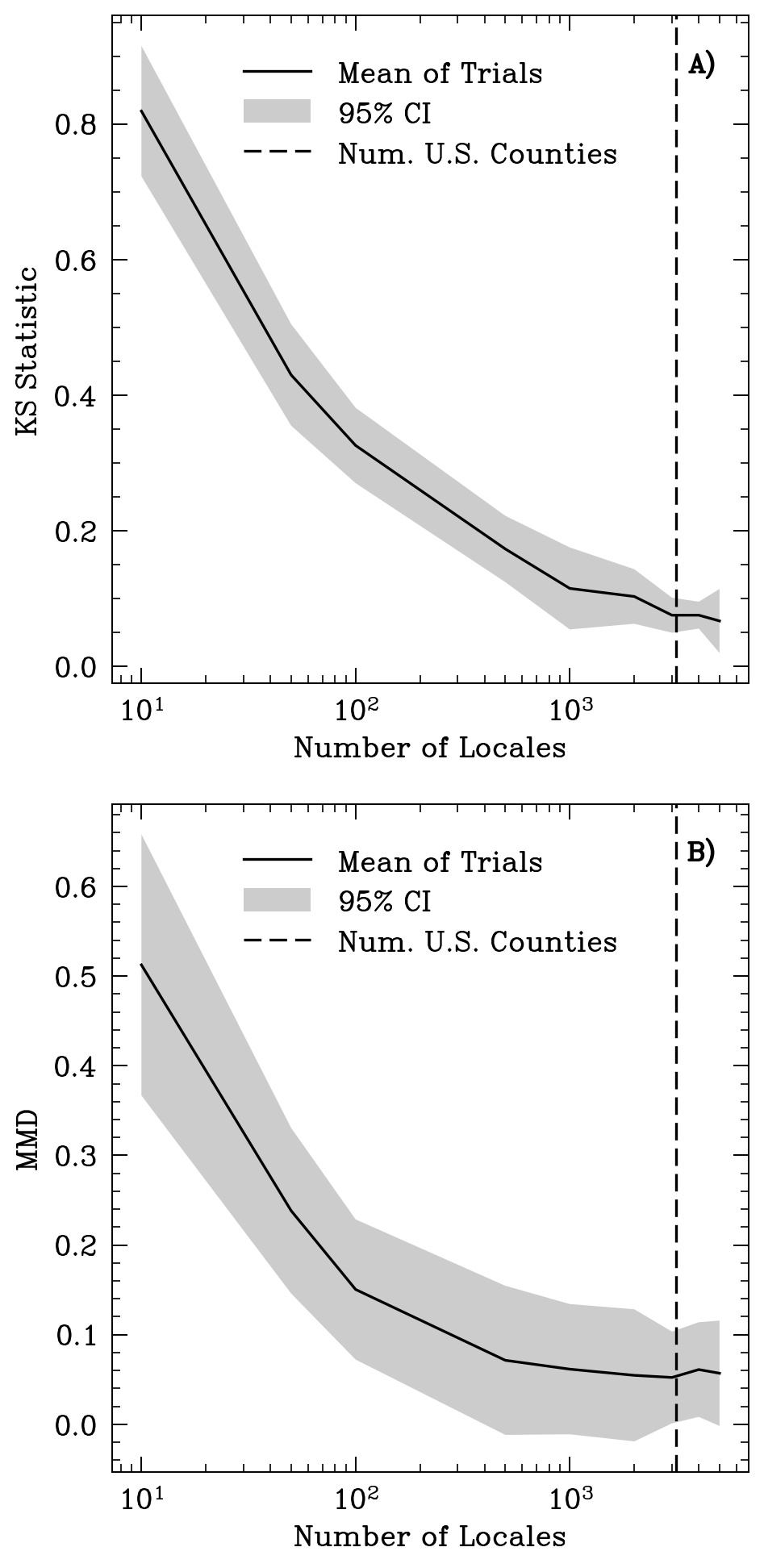}
    \caption{Convergence results. We consider $10$, $50$, $100$, $500$, $1000$, $2000$, $3000$, $4000$, $5000$ locales. We repeat each over $10$ trials to obtain confidence bands. For each trial, $100$ events are generated by the simulation. \textbf{A)} KS Statistic between the empirical CDF of uncensored events and the estimated CDF. \textbf{B)} For each trial, we draw $100$ samples from the estimated CDF. We show the empirical MMD between uncensored events and the drawn samples. We use a heuristic for the bandwidth parameter $\sigma$: the median of all absolute distances in the combined set of uncensored events and the drawn samples.}
    \label{fig:simresults}
\end{figure}

We wish to verify that, as locale size decreases, we are better able to reconstruct the true distribution of transportation event distances from their censored counterparts, and the (uncensored) samples we draw resemble their uncensored counterparts. We initialize the simulation environment with $1000$ locations and generate $100$ transportation events. We iteratively increase the number of locales (i.e. decrease their average sizes), mapping the events to their censored realizations, and investigate whether our approximation approaches the true distribution of transportation event distances.

In our trials, we see that the KS statistic between the empirical CDF of uncensored events and the estimated CDF reconstructed from their censored realizations decreases with locale count (Figure \ref{fig:simresults}A), as expected. We observe a similar effect with the empirical MMD between the uncensored events and the samples drawn from our reconstruction (Figure \ref{fig:simresults}B).

\begin{figure*}[t]
    \centering
    \includegraphics[width=0.85\textwidth]{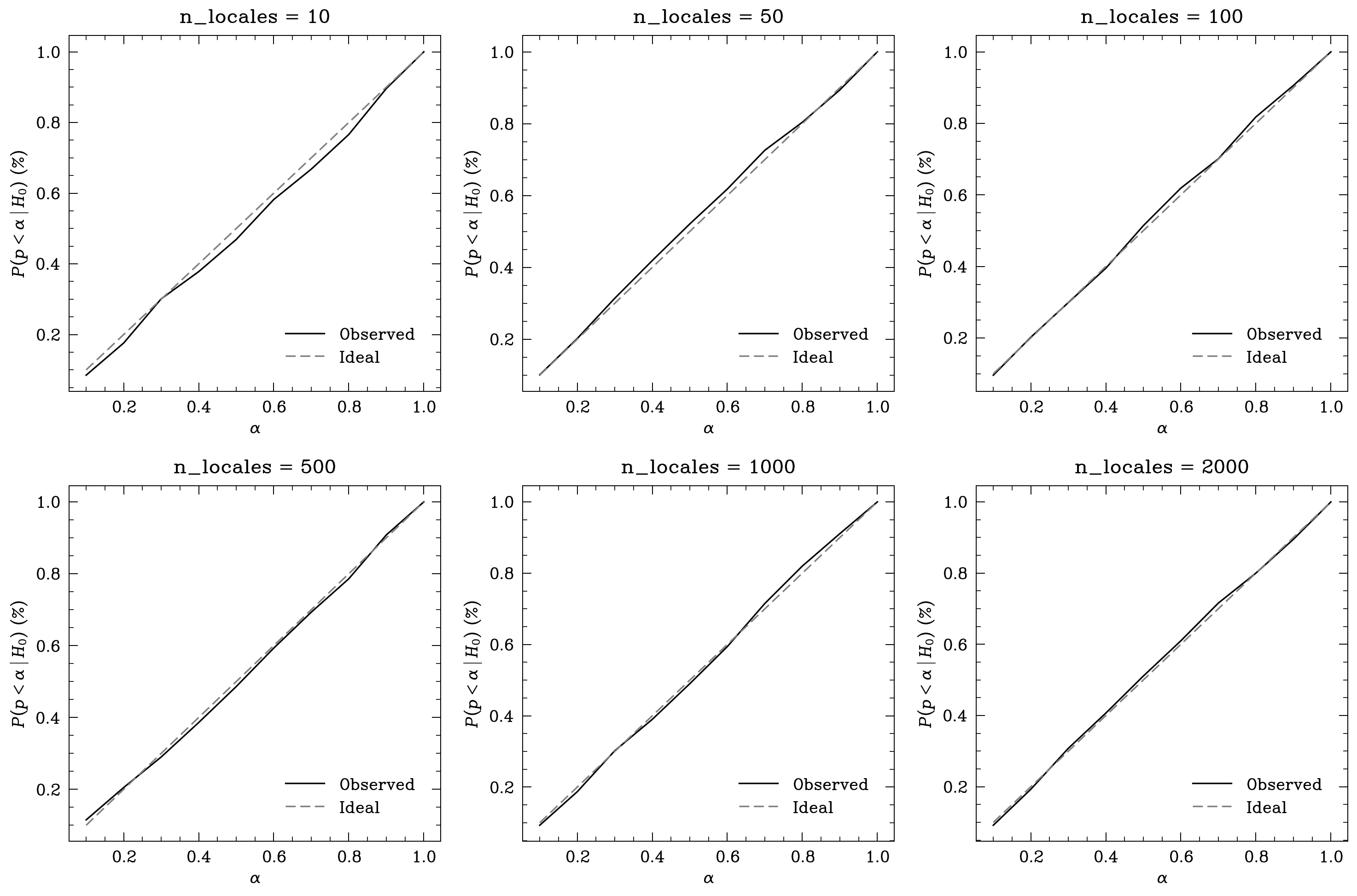}
    \caption{Monte Carlo U-test calibration results: maintenance of the Type I error rate. We consider the following numbers of locales: $10$, $50$, $100$, $500$, $1000$, $2000$. We initialize our simulation environment, generate $1000$ censored events, and run our Monte Carlo U-test with $E_A = E_B$, $m=n=100$, and $n_{\text{trials}}=100$. We run $1000$ hypothesis tests of this kind for each number of locales and illustrate the frequency of p-values below each significance level.}
    \label{fig:calibration_null}
\end{figure*}

We also empirically inspect the calibration of the Monte Carlo U-test (Algorithm \ref{alg:compare_events_repeated}). Using the same simulation environment, we vary the number of locales and confirm maintenance of the Type I error rate (Figure \ref{fig:calibration_null}) and uniformity of p-values (Table I)
under the null hypothesis.

\begin{table}[t]\label{tab:uniformity}
\centering
\begin{threeparttable}
\caption{Monte Carlo U-test calibration results: uniformity of p-values.\tnote{1}}
\begin{tabular}{|c|c|c|c|}
\hline
\textbf{Number of Locales} & \textbf{$\boldsymbol{\chi}^2$} & \textbf{P-Value} & \textbf{Significant}\tnote{2} \\ \hline
10                & 22.0        & 0.0167            & False                \\
50                & 14.0        & 0.1153            & False                \\
100               & 18.0        & 0.4190            & False                \\
500               & 18.0        & 0.3412            & False                \\
1000              & 18.0        & 0.4985            & False                \\
2000              & 22.0        & 0.4365            & False                \\ 
\hline
\end{tabular}
\begin{tablenotes}
\item[1] P-values from Figure \ref{fig:calibration_null} are distributed into $20$ equal-width bins, each bin covering a range of $5\%$. We report the results of chi-squared tests of uniformity.
\item[2] Indicates whether the corresponding p-value is deemed statistically significant (suggesting non-uniformity) after applying the Holm-Bonferroni adjustment to control family-wise error rate \cite{holm1979simple}.
\end{tablenotes}
\end{threeparttable}
\end{table}

\subsection{A Partial Re-Analysis of Maheswaran et al.}\label{sec:maheswaran}

\begin{table}[t]
\centering
\begin{threeparttable}
\caption{Breast cancer screening attendance by distance.\tnote{3}}
\begin{tabular}{|c|c|c|c|}
\hline
\textbf{Distance by category (km)} & \textbf{Invited} & \textbf{Attended} \\
\hline
$\geq8$ & 4641 & 3575 \\
$6$–$<8$ & 4982 & 3880 \\
$4$–$<6$ & 7871 & 6088 \\
$2$–$<4$ & 8068 & 6318 \\
$<2$ & 9306 & 7429 \\
\hline
\end{tabular}\begin{tablenotes}
\item[3] Reproduced from \textit{Socioeconomic deprivation, travel distance, location of service, and uptake of breast cancer screening in North Derbyshire, UK} by Ravi Maheswaran, Tim Pearson, Hannah Jordan, and David Black, (volume 60, page 210, copyright notice year 2006), with permission from BMJ Publishing Group Ltd. and the corresponding author \cite{maheswaran2006socioeconomic}.
\end{tablenotes}
\label{tab:attendance_summary}
\end{threeparttable}
\end{table}

An analysis of $1998$-$2001$ public health records from North Derbyshire, United Kingdom studied breast cancer screening uptake among women aged $50-64$ and its relationship with socioeconomic status and distance from screening facility. The authors had access to precise locations of the screening facilities, but only postal codes for resident addresses. The authors consider the minimum and maximum distances the residents could have traveled to each location within their postal code. When treating the censored transportation events as categorical data (Table \ref{tab:attendance_summary}), they conduct a chi-squared test to analyze the impact of distance on attendance \cite{maheswaran2006socioeconomic}.

Although this is a sensible analysis, there are limitations. Because it treats distance as a categorical variable, the ordinality of distance is ignored. Additionally, the chi-squared test aims to assess whether the expected frequency of events occurring in the distance bins differs between the ``Invited" and ``Attended" groups; even if the result is deemed significant, it does not specify the nature of the relationship. The authors also interpret distance as a continuous variable by mapping postal codes to their corresponding centroid location, a technique not uncommon in this situation, but we believe this mapping makes results difficult to interpret.

We apply Algorithm \ref{alg:compare_events_repeated} to the data in Table \ref{tab:attendance_summary}, with the alternative hypothesis that the distribution of distances from screening center for women invited for screening stochastically dominates the distribution of distances for women actually attending screening. We assume a comfortable $100$ km upper bound on distance traveled, run $100$ trials with $100$ samples drawn per trial, and do not find a statistically significant result ($p=0.8813$). We repeat the procedure for ``Not Attended" (i.e., those invited, but did not attend) and ``Attended," also failing to find a statistically significant result ($p=0.7671$).

\section{Conclusion}

In this work, we discuss the analysis of censored transportation event records. We leverage ideas from survival analysis to perform this estimation and suggest a sampling-based stochastic dominance test for comparing sets of censored transportation records. Future work should refine the approach and consider other tests for comparing survival curves, such as the Mantel–Cox test \cite{mantel1966evaluation}.

\section*{Acknowledgment}

We are grateful to Naoki Masuda, Mike Lujan, and Ericson Davis for helpful discussions. We also thank the developers of \textit{lifelines}, \textit{NetworkX}, \textit{SciPy}, and \textit{NumPy} for releasing their software on an open-source basis \cite{Davidson-Pilon2019, hagberg2008exploring, 2020SciPy-NMeth, harris2020array}. Finally, we thank OpenAI for developing ChatGPT-4, which was used to assist with code-writing supporting Figures $1-3$ and language refinement in the editing process.

\printbibliography

\end{document}